\documentclass[12pt,a4paper]{article}

\newcommand{\be}{\begin{equation}}                               %
\newcommand{\ee}{\end{equation}}                                 %
\newcommand{\bea}{\begin{eqnarray}}                              %
\newcommand{\eea}{\end{eqnarray}}                                %
\def\ad{{\rm ad\, }}                    %
\def\G{{\cal G}}                        %
\def\M{{\cal M}}                        %
\def\H{{\cal H}}                        %
\def\D{{\cal D}}                        %
\def\K{{\cal K}}                        %
\def\I{{\cal I}}                        %
\def\cR{{\cal R}}                       %
\def\R{{\mathbf R}}                     %                                        
\def\CG{{\check G}}                     %
\def\cG{{\check {\cal G}}}              %
\def\half{\frac{1}{2}}                  %
\begin{document}

\thispagestyle{empty}
\vspace*{0.4cm}
 
\begin{center}

{\Large\bf  Dynamical $r$-matrices  and Poisson-Lie symmetries \\
in the  chiral WZNW 
model\footnote{Talk presented at the Workshop on Integrable Theories, 
Solitons and Duality, July 2002, Sao Paulo, Brazil.}}

\end{center}

\vspace{0.8cm}

\begin{center}

L\'aszl\'o Feh\'er  \\

\medskip

Department of Theoretical Physics,  University of Szeged \\
Tisza Lajos krt 84-86, H-6720 Szeged, Hungary \\
e-mail: lfeher@sol.cc.u-szeged.hu

\end{center}

\vspace{0.8cm}

\begin{abstract}  

We briefly review the possible Poisson structures on the chiral 
WZNW phase space and discuss the associated Poisson-Lie groupoids.
Many interesting dynamical $r$-matrices appear naturally in this 
framework. Particular attention is paid to the special cases in 
which these $r$-matrices satisfy the classical dynamical Yang-Baxter 
equation or its Poisson-Lie variant.
 
\end{abstract}

\newpage

\section{Introduction}
\setcounter{equation}{0}

Let me start by recalling the standard classical 
dynamical Yang-Baxter  equation (CDYBE),
\begin{equation}
\big[r_{12}(\lambda),r_{23}(\lambda)\big]
+H_1^i \frac{\partial}{\partial \lambda^i}  r_{23}(\lambda)
+ \hbox{cycl. perm.}=0,
\label{1.1}\end{equation}
where $r(\lambda)\in \G\otimes \G$ and 
the variable $\lambda=\lambda^i H_i$ lies in a Cartan subalgebra 
of a simple Lie algebra.
This equation is the classical limit of the
Gervais-Neveu-Felder equation
\begin{equation}
R_{12}(\lambda + \hbar H_3) R_{13}(\lambda) R_{23}(\lambda + \hbar H_1)
=R_{23}(\lambda) R_{13}(\lambda + \hbar H_2) R_{12}(\lambda).
\label{1.2}\end{equation}
These equations govern the classical and quantum exchange
algebras of the chiral Bloch waves in the conformal Toda
and WZNW field theories on the cylinder \cite{GN,CG,BDF}.
They also appear in the description of the conformal blocks
of the WZNW model on the torus \cite{Feld} and 
in the study of Calogero-Moser models \cite{ABB}.
The CDYBE and its quantized version have interesting 
generalizations that play an important role
in quantum algebra and in the theory of integrable systems \cite{ES}. 

The generalizations of (\ref{1.1}) introduced 
by Etingof and Varchenko \cite{EV} 
result by replacing the Cartan subalgebra  by (the dual space of) 
any subalgebra $\H$ of any Lie algebra $\G$.  
Of course, one can also consider the spectral parameter dependent variant of 
these equations.
The most important special case, related to affine Kac-Moody algebras, 
is when the `dynamical variable' $\lambda$  belongs to the 
fixed point set of a Coxeter automorphism of a simple Lie algebra. 

In the context of the classical 
WZNW model, the variable $\lambda$ in (\ref{1.1})
is the logarithm of the monodromy of the chiral WZNW field.
Motivated by our longstanding interest in the WZNW 
model as well as by the intense current 
research activity around the CDYBE,
with J. Balog and L. Palla \cite{BFP}
we have recently explored the  most general Poisson structure 
that arises on the chiral WZNW phase 
space {\em without imposing the constraint that
the monodromy belongs to a maximal torus}. 
It turned out that these Poisson structures are parametrized by solutions 
of a generalization of the CDYBE, which we call the $G$-CDYBE.
We can actually find all solutions of the $G$-CDYBE as part of our
analysis of the WZNW model, which is meant to be a continuation 
of the pioneering papers \cite{G,FG}. 
Among the resulting dynamical $r$-matrices there is a particularly 
interesting class associated with Poisson-Lie symmetries acting on the
chiral WZNW phase space.  
When the Poisson-Lie symmetry degenerates into usual symmetry,  
then these $r$-matrices reduce to a canonical solution of the
CDYBE on $\G$ in the  sense of \cite{EV}, which
upon further Dirac reduction to suitable subalgebras of $\G$ (and by 
certain limiting procedures) reproduces many of the known $r$-matrices
of the Etingof-Varchenko type. 

Here I  present a brief review of the main results that we 
obtained in \cite{BFP}, touching also on   
their further clarifications published in \cite{PF,FM} .
For lack of time, 
I cannot deal with several related questions 
elaborated in the papers \cite{BFPinJPA,BFPinPLA,FGP,FehGab,FP}.
More detailed reviews of some aspects of  
our work can be found in \cite{Montreal,Dubna}.

\section{$G$-CDYBE from the chiral WZNW phase space}
\setcounter{equation}{0}

The WZNW model \cite{Wi} as a classical field theory on the cylinder
can be defined for any (real or complex) Lie group $G$
whose Lie algebra $\G$ is self-dual in the sense that it is  equipped
with an invariant, symmetric, non-degenerate 
bilinear form $\langle\ ,\ \rangle$.
The solution of the classical field equation for the $G$-valued
WZNW field, which is $2\pi$-periodic in the space variable, is
given by the product of left- and right-moving chiral WZNW fields 
that are quasi-periodic.
By restricting the `monodromy matrix' to lie in some {\em open 
submanifold} $\CG\subseteq G$,
we thus obtain the chiral WZNW phase space
\begin{equation}
\M_\CG:= \{  g \in C^\infty({\R}, G)  \,\vert\,
g(x + 2\pi) = g(x) M \quad  M\in \CG\}.
\label{2.1}\end{equation}
If one wishes to induce the Poisson structure of the full WZNW model from
Poisson brackets 
(PBs) of chiral fields varying in $\M_\CG$, 
then the only possibility
is to equip $\M_\CG$ with a PB of the following form:
\begin{equation}
\kappa \left\{g(x)\stackrel{\otimes}{,} g(y)\right\}_{WZ}^r
=\left(g(x)\otimes
g(y)\right)\left(
\half C \,{\rm sign}\,(y-x) + r(M)  
\right), \quad 0< x,y<2\pi.
\label{2.2}\end{equation}  
Here the interesting object is the `exchange r-matrix'
$r(M)= r^{ab}(M) T_a \otimes T_b \in \G\wedge \G$;
$C=T_a \otimes T^a$ where $\{ T_a\}$ and $\{T^a\}$ 
denote dual bases of $\G$, $\langle T_a, T^b\rangle =\delta_a^b$.
One way to derive this PB is to invert the 
symplectic form $\Omega^\rho_{WZ}$
on $\M_\CG$ found by Gawedzki \cite{G}:
\begin{equation}
 \frac{1}{\kappa}\Omega^\rho_{WZ}(g)=
- \frac{1}{2}\int_0^{2\pi}dx\,
 \langle (g^{-1}dg)  \stackrel{\wedge}{,}  (g^{-1}dg )'\rangle
-\frac{1}{2} \langle (g^{-1} dg)(0)
\stackrel{\wedge}{,} dM {\scriptstyle\,} M^{-1}\rangle + \rho(M)
\label{2.3}\end{equation} 
with some 2-form $\rho$ on $\CG$.
Another method that leads directly to (\ref{2.2}) uses the requirements that 
$J:= \kappa g' g^{-1}$ must be an affine Kac-Moody current with respect to
which $g$ is a chiral primary field, and that the WZNW solution space must
be obtained from the product of two independent chiral phase spaces by
imposing first class constraints. 
Both methods are explained in detail in \cite{BFP,Montreal}.
In either way, one can show that the PBs in (\ref{2.2}) are accompanied by 
\be
\kappa \{ g(x)\stackrel{\otimes}{,} M\}_{WZ}^r = 
\left(g(x)\otimes M\right)  \Theta(M)
\quad\hbox{and}\quad 
\kappa \{ M\stackrel{\otimes}{,} M\}_{WZ}^r =
\left( M\otimes M \right) \Delta(M),
\label{2.4}\ee
with 
\be
\Theta(M)=  r^+(M) - 
M_2^{-1}  r^-(M)M_2,
\quad 
\Delta(M)= \Theta(M) - M_1^{-1} \Theta(M) M_1
\label{2.5}\ee
where  $r^\pm := r \pm \frac{1}{2} C$, $M_1= M\otimes 1$, 
$M_2=1\otimes M$.
In fact \cite{BFP},    
{\em the Jacobi identity\footnote{As explained in \cite{BFPinPLA}, 
by setting $r(M)=0$ in (\ref{2.2}) one
obtains a quasi-Poisson structure \cite{AKS} on $\M_G$.}  
of the PB (\ref{2.2}) 
is equivalent to the following equation}:
\begin{equation}
\left[ r_{12}(M), r_{23}(M)\right]
+T_1^a \left( \frac{1}{2} \D_a^+ + 
r_a^{{\phantom{a}}b}(M) \D_b^-\right)  r_{23}(M)
+ \hbox{cycl. perm.}=
-\frac{1}{4} \hat f.
\label{2.6}\end{equation}
Here $\hat f:= f_{a b}^{\phantom{ab}c} T^a \otimes T^b \otimes T_c$ with 
$[T_a, T_b]=f_{ab}^{\phantom{ab}c}T_c$,  
$r_{23} = r^{ab}  (1\otimes T_a\otimes T_b)$ and 
$T^a_1= T^a \otimes 1 \otimes 1$ as usual;  
for any  function $\psi$ on $G$ we use   
\begin{equation}
\D_a^{\pm} = \cR_a \pm {\cal L}_a 
\quad\hbox{with}\quad 
({\cal R}_a \psi)(M):= \frac{d}{d t}
 \psi(Me^{t T_a} )\Big\vert_{t=0},
\quad 
({\cal L}_a \psi)(M):= \frac{d}{d t}
 \psi(e^{t T_a} M)\Big\vert_{t=0}.
\label{2.7}\end{equation} 
We call equation (\ref{2.6}) the $G$-CDYBE since it is a  
generalization of the CYBE for an  
$r$-matrix depending on a $G$-valued `dynamical variable'.
The $G$-CDYBE  becomes the standard modified-CYBE  
if $r$ is an $M$-independent constant.

In the symplectic formalism \cite{G,FG,Montreal} based 
on $\Omega^\rho_{WZ}$ (\ref{2.3}) 
the need to restrict to some submanifold of $G$ 
arises since the condition $d \Omega^\rho_{WZ}=0$ 
translates \cite{G} into 
\be
d\rho = \frac{1}{2} \langle [ M^{-1} dM , M^{-1} dM], M^{-1} dM \rangle, 
\label{2.8}\ee
which in general does not admit a global 
solution on $G$, as is well known.
In the alternative approach in which $(\M_\CG, \{\ ,\ \}_{WZ}^r)$ 
is required to be a (not necessarily 
non-degenerate) Poisson space,  the only condition is that 
$r: \CG \rightarrow \G\wedge \G$ must 
be a regular (smooth or holomorphic) 
solution of the $G$-CDYBE (\ref{2.6}) 
on some open submanifold $\CG\subseteq G$. 

Note that the exchange $r$-matrix can be constant, 
which is the case considered mostly in the early papers 
(see e.g. \cite{Bab,Blok,Fad,AS,FG} and references therein), 
if (\ref{2.6}) has such a solution.
The real forms of the simple Lie algebras that admit a 
constant solution \cite{Cahen} include 
the split real form, but exclude the compact one.

\section{The canonical $r$-matrix and its Dirac reductions}
\setcounter{equation}{0}

We actually have all solutions of the $G$-CDYBE 
locally in a neighbourhood of $e\in G$.
More precisely, we have an explicit one-to-one 
correspondence between the solutions
$\rho$ of (\ref{2.8}) and the 
solutions $r$ of (\ref{2.6}) around $e\in G$.  
Such a correspondence was derived originally by inverting 
$\Omega^\rho_{WZ}$ \cite{BFP},
but later a purely finite dimensional proof of it has 
also been given \cite{FehGab}.

A particularly interesting exchange $r$-matrix arises if the PB (\ref{2.2}) 
permits the `gauge action' of the group $G$ on $\M_\CG$, given by
\be
\M_{\CG}\times G \ni (g(x), h)\mapsto g(x) h,
\label{3.1}\ee
to operate as a classical $G$-symmetry generated by the 
logarithm of the monodromy matrix.
In this case the PB (\ref{2.2}) does not change 
under (\ref{3.1}) and we have
\begin{equation}
\kappa \{ g(x), m_a \}_{WZ}^r = g(x) T_a,
\qquad
\kappa \{ m_a, m_b\}_{WZ}^r =- 
f_{ab}^{\phantom{ab}c} m_c
\quad\hbox{for}\quad 
M= e^m,
\label{3.2}\end{equation}  
where $m=\log M$ lies in a neighbourhood of zero, $\cG \subset \G$, 
diffeomorphic to $\CG$ by the exponential map.
In the notion of `classical symmetry' it is understood that the 
elements of the symmetry group have zero PBs with everything.     
Let $r^0(m)$ denote an exchange $r$-matrix that permits such a
symmetry on $\M_\CG$.  
Upon comparing (\ref{2.4}), (\ref{2.5}) with (\ref{3.2}), it follows that  
(\ref{2.6}) can now be rewritten in the form 
\be
\left[r^0_{12}(m),r^0_{23}(m)\right]
+T_1^a \frac{\partial}{\partial m^a}  r_{23}^0(m)
+ \hbox{cycl. perm.}=
-\frac{1}{4} \hat f,
\label{3.3}\ee
and it is also easy to see that $r^0(m)$ must be $G$-equivariant, 
\be
r^0 (h m h^{-1})  = 
(h\otimes h) r^0(m) (h^{-1} \otimes h^{-1}) 
\qquad
\forall h\in G. 
\label{3.4}\ee
It is natural to search for an equivariant $r$-matrix by using the ansatz
\be 
r^0(m) = \langle T_a, f_0(\ad m) T_b \rangle T^a \otimes T^b,
\label{3.5}\ee
where $f_0(z)$ is assumed to be a holomorphic, odd complex function 
in a neighbourhood of zero on the complex plane.
Then (\ref{3.3}) yields a functional equation \cite{PF} for 
the holomorphic function $f_0$,
whose {\em unique} solution is provided by 
\be
f_0(z) = \frac{1}{2}\coth \frac{z}{2} - \frac{1}{z}.
\label{3.6}\ee
This joint solution of (\ref{2.6}) and (\ref{3.3}) has 
been derived in \cite{BFP} by means
of inverting the symplectic form (\ref{2.3}) for the 2-form 
\be
\rho_0(M)= -\frac{1}{2} \int_0^{2\pi} dx 
\langle d \bar m \stackrel{\wedge}{,} de^{x\bar m} \, 
e^{-x\bar m}\rangle,
\qquad
\bar m:= \frac{1}{2\pi}\log M, 
\label{3.7}\ee
which satisfies (\ref{2.8}) and renders $\Omega^{\rho_0}_{WZ}$  
invariant under (\ref{3.1}). 

Note that equation (\ref{3.3}) with (\ref{3.4}) is the
CDYBE on the Lie algebra $\G$ in the sense of \cite{EV} 
(or `modified' CDYBE on account 
of the non-zero right-hand side).
Somewhat implicitly, the solution of the CDYBE on a simple Lie algebra 
given by (\ref{3.5}), (\ref{3.6}) is already contained in \cite{EV}.
It was also found in \cite{AM} in the context of equivariant cohomology. 
Its uniqueness property under the ansatz (\ref{3.5}) 
is proven in \cite{PF}, where this $r$-matrix is called `canonical'.
(This uniqueness property should be compared with the description 
of the `moduli space' of the solutions of (\ref{3.3}), (\ref{3.4}) 
given in \cite{ES+}.)

The canonical $r$-matrix described above plays  a distinguished 
role among the solutions 
of the CDYBE in the Etingof-Varchenko sense.
To explain this, let $\H\subset \G$ be a self-dual subalgebra 
(on which $\langle\ ,\ \rangle$ remains non-degenerate), and consider 
the associated decomposition $\G= \H \oplus \H^\perp$. 
Then define $r^*: \check \H \rightarrow \mathrm{End}(\G)$ by 
\begin{equation}
r^*(\lambda)(X) =f_0(\ad \lambda)(X)\quad \forall X\in \H,
\qquad
r^*(\lambda)(Y)=
\frac{1}{2}\coth(\frac{1}{2}\ad\lambda)(Y)\quad
\forall Y\in \H^\perp.
\label{3.8}\end{equation}
We here use the Laurent series expansion 
of $\frac{1}{2}\coth(\frac{z}{2})$ 
around $z=0$, the $z^{-1}$ term in the expansion corresponds to
the operator $(\ad\lambda)^{-1}$ on $\H^\perp$.
The open domain $\check \H\subset \H$ 
is restricted by the 
condition\footnote{A non-empty domain exists, for example,
if $\H$ is a reductive subalgebra of a simple Lie algebra.} 
that $r^*(\lambda)$ must be well defined 
by formula (\ref{3.8}) for $\lambda\in\check\H$.
By using the identification $\mathrm{End}(\G)\simeq \G\otimes \G$
defined by the scalar product on $\G$,
it can be shown that $r^*$ solves the CDYBE on $\H\subset \G$:
\be
\left[r^*_{12}(\lambda),r^*_{23}(\lambda)\right]
+H_1^i \frac{\partial}{\partial \lambda^i}  r_{23}^*(\lambda)
+ \hbox{cycl. perm.}=
-\frac{1}{4} \hat f,
\qquad
\lambda\in \check\H\subset \H,
\label{3.9}\ee
where $H^i$ denotes a basis of $\H$.

The standard solution of the original CDYBE (\ref{1.1}),
which first appeared in \cite{BDF} (see also \cite{Chu5,BBT}),  
is recovered from (\ref{3.8}) by 
taking $r:= r^* \pm \frac{1}{2} C$ and identifying   
$\H$ with a Cartan subalgebra of a simple Lie algebra. 

The passage from $r^0$ to $r^*$ corresponds to Dirac reduction 
in two senses. First, the phase space $\M_\CG$ equipped with
the canonical exchange $r$-matrix can be reduced by restricting
the monodromy to $\exp(\check\H)$, whereby the Dirac brackets 
of the chiral WZNW field take the same form as 
the PB in (\ref{2.2}), but with $r^*$
appearing in the role of the exchange $r$-matrix \cite{Dubna}.
Second, as developed in \cite{FGP}, 
 the Dirac reduction can also be implemented on
the Poisson-Lie groupoids that enter the geometric interpretation
of the CDYBE introduced in \cite{EV}.

Finally, it is worth noting that formula (\ref{3.8}) contains
Felder's celebrated 
spectral parameter dependent elliptic 
dynamical $r$-matrices \cite{Feld}
and some generalizations of them, too.
They are obtained  \cite{EV,ES3,FP}
by taking $\G$ to be an affine Kac-Moody type 
Lie algebra with $\H$ being a grade zero subalgebra in an integral   
gradation, and applying an evaluation homomorphism.  

\section{Exchange $r$-matrix compatible  with any PL structure}
\setcounter{equation}{0}
 
We have seen that by an appropriate choice of the exchange $r$-matrix
the gauge action (\ref{3.1}) can be interpreted as a classical $G$-symmetry 
on $\M_\CG$. 
Interestingly, we can achieve the same with respect to any 
(coboundary) Poisson-Lie (PL) structure on $G$.

Let us equip the group $G=\{ h\} $ with a PL structure 
by means of the Sklyanin bracket 
\be
\kappa \{ h \stackrel{\otimes}{,} h\}_{R^\nu} = [h\otimes h, R^\nu],
\label{4.1}\ee
where $R^\nu\in \G\wedge \G$
 is a {\em constant} $r$-matrix satisfying    
\be
[R_{12}^\nu, R_{23}^\nu] +\hbox{cycl. perm.} = -\nu^2 \hat f
\label{4.2}\ee
for some constant $\nu$.
(Note that $\nu$ must be purely imaginary, or zero, if $G$ is compact.)
Then look for the conditions on $r(M)$ that
guarantee the standard right action (\ref{3.1}) 
of $G$ on $\M_\CG$  to be a PL action. 
In fact, we find the requirement   
\be
K(hMh^{-1}) = (h\otimes h) 
K(M) (h^{-1}\otimes h^{-1})
\quad\hbox{for}\quad
K(M):= r(M) -R^\nu.
\label{4.3}\ee
This means that {\em the gauge action (\ref{3.1}) of 
$(G, \{\ ,\ \}_{R^\nu})$ on $(\M_\CG, \{\ ,\ \}_{WZ}^r)$ is a PL symmetry  
if and only if  
the exchange r-matrix $r(M)$ is such a solution of
(\ref{2.6}) for which the difference $(r(M)- R^\nu)$ is $G$-equivariant.} 

By using the substitution $r(M)=R^\nu+ K(M)$ together with (\ref{4.2}) 
and the equivariance condition (\ref{4.3}), the $G$-CDYBE (\ref{2.6}) 
can be rewritten in the form   
\begin{equation}
\left[ K_{12}(M), K_{23}(M)\right]
-\frac{1}{2} T_1^a  \D_a^+ 
K_{23}(M)
+ \hbox{cycl. perm.}=
(\frac{1}{4}-\nu^2) \hat f.
\label{4.4}\end{equation}
This equation for an equivariant $K: \CG\rightarrow \G\wedge\G$
may be referred to as the {\em PL-CDYBE on $G$} since it 
guarantees PL $G$-symmetry on the chiral WZNW phase space. 
It is remarkable that the reference $r$-matrix $R^\nu$ enters into 
this equation only through the constant $\nu$ in (\ref{4.2}). 

In a neighbourhood of $e\in G$, 
it is natural to search for  $K(M)$ with the aid of the ansatz
\be 
K(M) = \langle T_a, f_\nu(\ad m) T_b \rangle T^a \otimes T^b,
\qquad
m=\log M,
\label{4.5}\ee
where $f_\nu(z)$ is assumed to be a holomorphic, odd complex function 
in a neighbourhood of zero on the complex plane.
Then (\ref{4.4}) yields a functional equation \cite{FM} 
for the holomorphic function $f_\nu$,
whose {\em unique} solution is found to be  
\be
f_\nu(z) = \frac{1}{2}\coth \frac{z}{2} - \nu\coth \nu z.
\label{4.6}\ee
The exchange $r$-matrices 
provided by this result were first found in \cite{BFP} by 
using a different (more complicated) method,
their uniqueness under the ansatz (\ref{4.5}) has been established 
in \cite{FM}.
 
Some further remarks are here in order.
First, note that for $\nu=0$ $f_\nu$ in (\ref{4.6}) becomes 
the function $f_0$ in (\ref{3.6}).
The corresponding exchange $r$-matrix is thus compatible
with classical $G$-symmetry (for $R^0=0$) as well as with PL symmetry
for any antisymmetric solution $R^0\neq 0$ of the CYBE. 
Second, if  $\nu =\frac{1}{2}$ then $r=R^{\frac{1}{2}}$,
which is the case of the constant exchange $r$-matrices.
Third,  as mentioned before, for a compact Lie algebra $\G$
constant exchange $r$-matrices do not exist,  
because of the negative sign on the right-hand side of (\ref{2.6}),
but the above solutions of (\ref{2.6}) are available 
also in this case using a purely imaginary $\nu$ in (\ref{4.2}).

\section{Finite dimensional phase spaces related to WZNW}
\setcounter{equation}{0} 

The chiral WZNW Poisson structure (\ref{2.2}) is fixed once
a solution of the $G$-CDYBE (\ref{2.6}) is given. 
It appears interesting that the WZNW exchange $r$-matrices 
also encode the PBs on certain finite dimensional Poisson manifolds.
Indeed, it has been found in \cite{BFP} that 
{\em on the manifold
\be
P:= \check G \times G \times \check G = \{ (M^L, g, M^R)\},
\label{5.1}\ee
the following formula defines  a PB, $\{\ ,\ \}_{P}^{r}$,  
for any solution $r: \CG \rightarrow \G\wedge\G$ of the $G$-CDYBE:}
\bea
&&
\kappa \{ g_1, g_2\}_P^r = g_1 g_2  r(M^R) -  r(M^L) g_1 g_2 
\nonumber\\
&& \kappa \{ g_1, M^R_2\}_P^r = g_1 M_2^R  \Theta(M^R)
\nonumber\\
&& \kappa \{ g_1, M_2^L\}_P^r = M_2^L  \Theta(M^L) g_1
\nonumber\\
&&\kappa \{ M^R_1, M^R_2\}_P^r = M^R_1 M^R_2 \Delta(M^R)
\nonumber\\
&&\kappa \{ M^L_1, M^L_2\}_P^r = - M^L_1 M^L_2 \Delta(M^L)
\nonumber\\
&&\kappa \{ M^R_1, M^L_2\}_P^r =0.
\label{5.2}
\eea
We use $\Theta$ and $\Delta$ defined in (\ref{2.5}), and to maintain 
the obvious similarity to (\ref{2.4}) we even included the arbitrary
constant $\kappa$ (the classical level parameter) into this definition.  
We stress that $g\in G$ is here $x$-independent. 
In fact, $(P, \{\ ,\ \}_P^r)$ is an example of a Poisson-Lie  
groupoid in the sense of \cite{We}.
This PL groupoid `extracted' from $(\M_\CG, \{\ ,\ \}_{WZ}^r)$
provides a geometric interpretation of the $G$-CDYBE analogous
to the interpretation of the Etingof-Varchenko $r$-matrices \cite{EV}.   
If $r$ is associated with classical $G$-symmetry on $\M_\CG$ as 
described in Section 3, 
then our PL groupoid is essentially identical with the 
`dynamical PL groupoid'
of \cite{EV} that encodes the CDYBE (\ref{3.3}) on $\G$.   

In analogy with the CDYBE on $\G$, 
the PL-CDYBE (\ref{4.4}) admits a `canonical' interpretation \cite{FM}
in terms of well known objects of PL geometry, which 
is nicer than the general case (\ref{5.2}).  
To describe this, let us now denote the elements of $P$ differently as
\be
P:= \CG \times G \times \CG = 
\{ (\Omega^L, g, \Omega^R)\,\vert\, \Omega^{L,R}\in \CG,\, g\in G \},
\label{5.3}\ee 
where $\check G \subset G$ is some open submanifold.
Then let $\cR:=R^{\frac{1}{2}}\in \G\wedge \G$ be a constant solution 
of (\ref{4.2}) with $\nu=\frac{1}{2}$ and let 
$\K: \CG \mapsto \G\wedge \G$ be a (smooth or holomorphic) map.
Now consider the following ansatz for a PB, $\{\ ,\ \}_{can}$, on $P$:
\bea
&&\{ g_1, g_2\}_{can}= (\cR+\K(\Omega^L)) g_1 g_2 
- g_1 g_2 (\cR+\K(\Omega^R)) 
\nonumber\\
&& \{ g_1, \Omega^R_2 \}_{can} = g_1 ( \cR^- \Omega^R_2 - 
\Omega^R_2 \cR^+)
\nonumber\\
&& \{ g_1, \Omega^L_2 \}_{can} = ( \cR^- \Omega^L_2 - 
\Omega^L_2 \cR^+) g_1
\nonumber\\
&& \{ \Omega^R_1, \Omega^R_2\}_{can}=
 - \cR \Omega^R_1 \Omega^R_2 - \Omega^R_1 \Omega^R_2 \cR  
 +\Omega^R_1 \cR^- \Omega^R_2 + 
\Omega^R_2 \cR^+ \Omega^R_1
\nonumber\\
&& \{ \Omega^L_1, \Omega^L_2\}_{can}= \cR \Omega^L_1 \Omega^L_2 +
\Omega^L_1 \Omega^L_2 \cR  - 
\Omega^L_1 \cR^- \Omega^L_2 - \Omega^L_2 \cR^+ \Omega^L_1
\nonumber\\
&&\{ \Omega^R_1, \Omega^L_2\}_{can}=0.
\label{5.4}\eea
Note that $\cR^{\pm}:= \cR \pm \frac{1}{2} C$ and that the
dynamical $r$-matrix $\K$ appears only in the first line 
of formula (\ref{5.4}).
We assume that $\K$ is a {\em $G$-equivariant} map, since anyhow this is 
required locally around $e\in G$ by the Jacobi identity
$\{\{ g_1, g_2\}_{can}, \Omega^L_3\}_{can} 
+ \hbox{cycl. perm}=0$ and its counterpart with $\Omega^R$.
The only nontrivial Jacobi identity to check is the one involving 
$\{\{g_1, g_2\}_{can}, g_3 \}_{can}$.
This condition is found to be {\em equivalent to the following 
version of the PL-CDYBE}: 
\be
[\K_{12}, \K_{23}] + \half T^a_1 \D_a^+ \K_{23} 
+\hbox{cycl. perm.} = {\I} \qquad\hbox{on}\qquad \check G,
\label{5.5}\ee
where $\cal I$ is an arbitrary $G$-invariant constant element 
of $\G\wedge \G\wedge \G$.

If we set $\K=-K$ and $\I=(\frac{1}{4}-\nu^2)\hat f$, then 
(\ref{5.5}) becomes identical to (\ref{4.4}).
If $K$ is given by (\ref{4.5}) with (\ref{4.6}), then 
the PB in (\ref{5.2}) can be converted 
into (a multiple of) the PB in (\ref{5.4}) by a certain change of variables.
This will be described in detail in a future publication.
  
In fact \cite{STS,AF}, for $\K=0$ the PB (\ref{5.4}) becomes  
the canonical PB of the Heisenberg double of the PL group $G$ equipped
with the Sklyanin PB that belongs to $\cR$, if one further sets 
$\Omega^R=g^{-1} \Omega^L g$.
Thus $\Omega^{L}$ and $\Omega^R$ define directly the momentum maps 
that generate the natural PL actions of $G$ on $(P, \{\ ,\ \}_{can})$ 
that act by left and right-multiplications on $g$.    

It is known \cite{STS,AF} how to quantize the Heisenberg double, i.e., the PB
(\ref{5.4}) with $\K=0$.
It is an interesting open problem to perform the quantization of $(P,\{\ ,\ \}_{can})$ 
in the case 
\be
\K(\Omega)=-\langle T_a, f_\nu (\ad (\log\Omega)) T_b \rangle T^a \otimes T^b 
\ee
with the function $f_\nu$ in (\ref{4.6}), which are the unique solutions of 
the PL-CDYBE (\ref{5.5}) given in terms of a complex analytic function. 
One should consider the quantization of the PB (\ref{5.2}) also in 
the general case; the resulting structure should be related to 
the `quantum algebraic properties' of the WZNW conformal field theory.   
This appears to be a natural idea in the context of the programme to 
canonically quantize the WZNW model and to study the associated 
`chiral zero modes' (see \cite{AF},
 \cite{Furlan} and references therein).   

\section{Concluding remarks}
\setcounter{equation}{0} 

The results reported so far in this  talk can be extended in several directions.
For example, one can consider the dynamical $r$-matrices that arise in 
the classical WZNW model with twisted boundary condition associated
with a finite order automorphism, $\mu$, of the group $G$.
Denote also by $\mu$ the induced automorphism of $\G$, of order $N$ say,
and suppose that it preserves the `scalar product' $\langle\ ,\ \rangle$ on $\G$. 
In this case the full WZNW field satisfies 
$g_{WZ}(\sigma+2\pi,\tau)=\mu(g_{WZ}(\sigma,\tau))$ and the corresponding 
chiral fields obey 
\be
g(x +2\pi) = \mu(g(x)) M,
\qquad
M\in G.
\label{6.1}\ee
Let us assume for simplicity that $G$ is a {\em complex simple} Lie group,
and use the decomposition 
\be
\G=\oplus_{a=0}^{N-1} \G_a,
\qquad
\mu(X)= \exp({2\pi i \frac{a}{N}}) X\quad \forall X\in\G_a.
\label{6.2}\ee
In order to obtain a twisted analogue of the canonical $r$-matrix, 
we restrict the monodromy to the form $M=e^{\lambda}$ with 
$\lambda$ varying in an open domain, $\cG_0$, in $\G_0$.
Then we  enquire about the PB of the 
$\mu$-twisted chiral WZNW field under the assumption that it enjoys 
classical $G_0$-symmetry.
In fact, we find that such a PB again has the form (\ref{2.2}) 
with the exchange $r$-matrix 
$r_\mu: \cG_0 \rightarrow \G\otimes \G \simeq {\mathrm{End}}(\G)$ given by
\begin{equation}
r_\mu(\lambda)=\left\{
\begin{array}{cc} 
f_0(\ad \lambda)  &\mbox{on $\G_0$}\\
\frac{1}{2}\coth\left(\frac{1}{2}\ad \lambda -i\pi \frac{a}{N}\right) 
 &\mbox{on $\G_a$ for $a\neq 0$},
\end{array}\right. 
\label{6.3}\end{equation}
where $f_0$ appears in (\ref{3.6}).
This dynamical $r$-matrix was originally 
found\footnote{In \cite{ES+} the simplicity of $\G$ is not 
assumed, see also \cite{FP,AMnew}.} 
 \cite{ES+} as a solution of the 
CDYBE on $\G_0$ (eq.~(\ref{3.9}) with $\H:=\G_0$).
It can be derived  by calculating the PBs on the $\mu$-twisted 
chiral WZNW phase space with $M\in \check G_0$ and the symplectic form
defined by using (\ref{2.3}) with the restriction of $\rho_0$ in (\ref{3.7}) to $\CG_0$. 
If $\G_0$ is non-Abelian, then further Dirac reduction to a Cartan subalgebra 
of $\G_0$ (and intermediate cases) is also possible, yielding `twisted analogues'    
of the $r$-matrices in (\ref{3.8}).
In the cases for which $\G_0$ is Abelian, 
the Wakimoto type free field realizations of the 
chiral WZNW field can be worked out following the lines of
\cite{BFPinJPA}.

So far I have performed the above analysis under the simplifying assumption
that $\G$ is complex and simple, but it should not be difficult to 
generalize it to any self-dual Lie algebra equipped with a finite
order automorphism compatible with the scalar product. 
 
Dynamical exchange $r$-matrices appear not
only in the (twisted) chiral WZNW model, but also in its intriguing 
generalization introduced recently by Klimcik.
In particular, Felder's $r$-matrix \cite{Feld} encodes the PBs of the chiral fields
in this model \cite{Klim}.
Therefore it is natural to expect that 
this model with twisted boundary condition should accommodate 
in its PBs the generalizations of Felder's $r$-matrices \cite{ES3,FP} 
associated with twisted affine Kac-Moody algebras, but further 
work is needed to clarify the situation. 
  
Another problem, which is currently under investigation, concerns the  
correspondence between some of the $r$-matrices mentioned in the talk 
and spin Calogero-Moser type integrable systems.  
At the classical level, 
it seems straightforward how to generalize the method of \cite{ABB,LiXu} 
to any solution of the CDYBE on $\H\subset \G$ (\ref{3.9}) as well as to
solutions of the spectral parameter dependent CDYBE on self-dual Lie algebras.
It is not clear however if this method will lead to new and interesting integrable 
systems or not. 
We also would like to see if the natural generalization 
of the CDYBE on $\G$ given by the PL-CDYBE (\ref{5.5}) 
is related to (perhaps spin Ruijsenaars type)
integrable systems. 
I hope to report on these questions on another occasion.
            
\bigskip\medskip
\noindent{\bf Acknowledgements.}
I wish to thank the organizers of the meeting at IFT for 
the invitation to Sao Paulo and for the superb hospitality.
This work was also supported by the Hungarian 
Scientific Research Fund (OTKA) under T034170, 
T029802, T030099, M028418  and M036804.


\begin{thebibliography}{99}

\bibitem{GN}
J.-L. Gervais and A. Neveu,  {\sl Nucl. Phys.} {\bf B 238} (1984) 125.

\bibitem{CG}
E. Cremmer and J.-L. Gervais, {\sl Commun. Math. Phys.} {\bf 134} (1990)  619.

\bibitem{BDF}
J. Balog, L. D\c{a}browski and L. Feh\'er, 
{\em Phys. Lett.} {\bf B 244} (1990) 227. 

\bibitem{Feld} 
G. Felder,
%%% Conformal field theory and integrable systems 
%%% associated to elliptic curves, 
pp.~1247-1255 in: Proc. ICM Z\"urich,   1994 
(Birkh\"auser,  1994), 
arXiv:hep-th/940715.

\bibitem{ABB} J. Avan, O. Babelon and E. Billey, 
{\sl Commun. Math. Phys.}  {\bf 178} (1996) 281, 
arXiv:hep-th/9505091.

\bibitem{ES} P. Etingof and O. Schiffmann,
Lectures on the dynamical Yang-Baxter equations, 
arXiv:math.QA/9908064.

\bibitem{EV} P. Etingof and A. Varchenko, 
{\sl Commun. Math. Phys.} {\bf 192} (1998) 77, 
arXiv:q-alg/9703040. 

\bibitem{BFP}
J. Balog, L. Feh\'er and L. Palla, 
{\sl Phys. Lett.} {\bf B 463} (1999) 83, 
arXiv:hep-th/9907050;\\
J. Balog, L. Feh\'er and L. Palla, 
{\sl Nucl. Phys.} {\bf B 568} (2000) 503, 
arXiv:hep-th/9910046.

\bibitem{G}
K. Gaw\c{e}dzki, {\sl Commun. Math. Phys.} {\bf 139} (1991) 201.

\bibitem{FG}
F. Falceto and K. Gaw\c{e}dzki, 
{\sl J. Geom. Phys.} {\bf 11} (1993) 251, 
arXiv:hep-th/9209076.

\bibitem{PF}
B.G. Pusztai and L. Feh\'er,
{\sl J. Phys. A}  {\bf 34} (2001) 10949, 
arXiv:math.QA/0109082.

\bibitem{FM}
L. Feh\'er and I. Marshall,
{\sl Lett. Math. Phys.} {\bf 62} (2002) 51,
arXiv:math.QA/0208159.

\bibitem{BFPinJPA}
J. Balog, L. Feh\'er and L. Palla, 
{\sl J. Phys. A} {\bf 33} (2000) 945, 
arXiv:hep-th/9910112.

\bibitem{BFPinPLA}
J. Balog, L. Feh\'er and L. Palla, 
{\sl Phys. Lett.} {\bf A 277} (2000) 107, 
arXiv:hep-th/0007045.

\bibitem{FGP}
L. Feh\'er, A. G\'abor and B.G. Pusztai,
{\sl J. Phys. A} {\bf 34} (2001) 7235, 
arXiv:math-ph/0105047.

\bibitem{FehGab}
L.\ Feh\'er and A. G\'abor,
%%%Interpretations and constructions of dynamical $r$-matrices,
pp.~331-336 in: {Quantum Theory and Symmetries}, eds. E. Kapuscik et al
(World Scientific, 2002), 
arXiv:hep-th/0111252. 

\bibitem{FP}
L. Feh\'er and B.G. Pusztai,
{\sl Nucl. Phys.} {\bf B 621} (2002) 622,
arXiv:math.QA/0109132.

\bibitem{Montreal}
J. Balog, L. Feh\' er and L. Palla,
%%% On the chiral WZNW phase space, exchange
%%% r-matrices and Poisson-Lie groupoids, 
pp.~1-19 in: CRM Proceedings and Lectures Notes, 
Volume 26, eds.~J. Harnad et al (AMS, 2000), 
arXiv:hep-th/9912173.

\bibitem{Dubna}
L. Feh\'er,
{\sl Phys.  Atom. Nucl.} {\bf 65} (2002) 1023, 
arXiv:math-ph/0104027. 

\bibitem{Wi}
E. Witten, {\sl Commun. Math. Phys.}  {\bf 92},   455 (1984).

\bibitem{AKS}
A. Alekseev,Y. Kosmann-Schwarzbach and  E. Meinrenken, 
{\sl Canad. J. Math.} {\bf 54} (2002) 3,
arXiv:math.DG/0006168. 

\bibitem{Bab}
O. Babelon, 
{\sl Phys. Lett.} {\bf B 215} (1988) 523.

\bibitem{Blok}
B. Blok, 
{\sl Phys. Lett.}  {\bf B 233} (1989) 359.

\bibitem{Fad}
L.D. Faddeev, 
{\sl Commun. Math. Phys.} {\bf 132} (1990) 131.

\bibitem{AS}
A. Alekseev and S. Shatashvili, 
{\sl Commun. Math. Phys.} {\bf 133} (1990) 353.

\bibitem{Cahen}
M. Cahen, S. Gutt and J. Rawnsley, 
{\sl Contemp. Math.} {\bf 179} (1994) 1.

\bibitem{AM} 
A. Alekseev and E. Meinrenken,
{\sl Invent. Math.} {\bf 139} (2000) 135, 
arXiv:math.DG/9903052.

\bibitem{ES+}
P. Etingof and O. Schiffmann, 
{\sl Math. Res. Lett.}  {\bf 8} (2001) 157, 
arXiv:math.QA/0005282.

\bibitem{Chu5}
M. Chu, P. Goddard, I. Halliday, D. Olive and A. Schwimmer,
{\sl Phys. Lett.} {\bf B 266} (1991) 71.

\bibitem{BBT}
O. Babelon, F. Toppan and L. Bonora, 
{\sl Commun. Math. Phys.} {\bf 140} (1991) 93.

\bibitem{ES3}
P. Etingof and O. Schiffmann, 
{\sl Math. Res. Lett.} {\bf 6} (1999) 593, 
arXiv:math.QA/9908115.

\bibitem{We}
A. Weinstein, {\sl J. Math. Soc. Japan} {\bf 40} (1988) 705. 

\bibitem{STS}
M.A. Semenov-Tian-Shansky, 
{\sl Publ. RIMS} {\bf 21} (1985) 1237;\\
M.A. Semenov-Tian-Shansky, {\sl Theor. Math. Phys.} {\bf 93} (1992), 1292,
arXiv:hep-th/9304042.

\bibitem{AF}
A. Alekseev  and L.D. Faddeev,
{\sl Commun. Math. Phys.} {\bf 141} (1991), 413;\\
A. Alekseev  and L.D. Faddeev,  An involution and dynamics for the
$q$-deformed quantum top, 
arXiv:hep-th/9406196.

\bibitem{Furlan}
P. Furlan, L.K. Hadjiivanov and I.T. Todorov,
Chiral zero modes of the SU(n) WZNW model, 
arXiv:hep-th/0211154.

\bibitem{AMnew}
A. Alekseev and E. Meinrenken,
Clifford algebras and the classical dynamical Yang-Baxter equation,
arXiv:math.RT/0209347.

\bibitem{Klim}
C. Klimcik, 
Quasitriangular WZW model, 
arXiv:hep-th/0103118.

\bibitem{LiXu}
L.C. Li and P. Xu, 
{\sl Commun. Math. Phys.} {\bf 231} (2002) 257,
arXiv:math.QA/0105162.

\end{thebibliography}
\end{document}